\documentclass[aip,graphicx]{revtex4-1}
\usepackage{graphicx}
\usepackage{amsmath}
\usepackage{amssymb}
\usepackage{xcolor}
\usepackage{subfigure}
\DeclareRobustCommand{\rchi}{{\mathpalette\irchi\relax}}
\newcommand{\irchi}[2]{\raisebox{\depth}{$#1\chi$}}

\begin{document}

\title{Spectrally tunable linear polarization rotation using stacked metallic metamaterials}

\author{Xavier Romain, Fadi I. Baida and Philippe Boyer}
\email[]{xavier.romain@femto-st.fr}
\affiliation{D\'epartement Optique - Institut FEMTO-ST UMR 6174, Universit\' e Bourgogne Franche-Comt\'e - CNRS, 25030 Besan\c con, France}

\date{\today}

\begin{abstract}
We theoretically study  the transmission properties of a stack of metallic metamaterials and show that is able to achieve a perfect transmission selectively exhibiting broadband (Q$<10$) or extremely narrowband (Q$>10^5$) polarization rotation. We especially highlight how the arrangement of the stacked structure, as well as the metamaterial unit cell geometry, can highly influence the transmission in the spectral domain. For this purpose, we use an extended analytical Jones formalism that allows us to get a rigorous and analytical expression of the transmission. Such versatile structures could find potential applications in polarimetry or in the control of the light polarization for THz waves.
\end{abstract}


\maketitle 

\section{Introduction}
Many efforts have been made to the physical understanding and to the theoretical modeling of enhanced transmission through periodically patterned metallic screens, ranging from the visible to the microwave domains \cite{art:Ebbesen98,art:Grady13,art:GarciaVidal10,art:Ruan06,art:Garcia07,art:Delgado13,art:Lalanne00,art:Boyer12,art:Baida11}. It is well-established that enhanced transmission is attributed to the excitation of cavity modes \cite{art:Astilean00,art:Baida04} or surface plasmons \cite{art:MartinMoreno01,art:Thio99,art:Poujet07}. The concept of extraordinary transmission has further been used in metallic multilayers \cite{art:Li07,art:Cheng08} to develop new optical functionalities, but it remains to be fully investigated for the terahertz (THz) and microwave domains \cite{art:Park09}. More recently, there has been an increased interest in polarization properties of stacked metamaterials \cite{art:Nader14,art:Marwat14}. Polarization rotation has been obtained using cascaded metallic hole arrays, featuring a relatively broad spectral bandwidth with quality factors $Q <10$ \cite{art:Li10,art:Zhang13,art:Xu13,art:Wei11,art:Wang16}. In the same manner, patch-based reflecting arrays were proposed to perform efficient polarization conversion covering a very broad spectral bandwidth \cite{art:Yang14,art:chen14,art:Zhang15,art:Yin15,art:Sui16} with $Q$ close to unity.

In this paper, we propose a stack of Periodic Metallic Polarizers (PMP) performing polarization-selective rotation of the incident electromagnetic field. Thus, we consider a rotation in such a way that only one component of the transverse electric field is transmitted and rotated by the structure. We demonstrate that stacked PMP patterned with subwavelength rectangular holes allow for extremely tunable spectral bands. Indeed, we study how such structures are able to achieve either a broadband Linear Polarization Rotation (LPR), with $Q < 10$, and how they are specially adapted to achieve extremely narrowband LPR, with $Q>10^5$, while ensuring a perfect transmission. Specifically, we propose a theoretical study for a number $N \geq 3$ of stacked PMP to investigate LPR in the THz regime. We show that total LPR is achieved thanks to the multiple reflections between the PMP.

In a previous paper, we have theoretically investigated the special case of a two-PMP stack in a polarizer analyzer configuration \cite{art:Romain16}. Therein, we have demonstrated that the multiple reflections between the two PMP leads to interesting transmission responses of the structure. Thanks to this phenomenon, we have shown that two stacked PMP can be used as an ultra-sensitive device for characterizing electro-optical materials in the THz region. It is, however, not adapted for realizing spectrally tunable LPR because a stack of at least three PMP is necessary for the design of LPR.

In section 2, we present a new theoretical model used to describe the LPR effects. Then, we give an analytical expression of the transmission to quantify the polarization properties of the structure. In section 3, we design the PMP geometry to selectively achieve optimized broad or narrow spectral bandwidths. We point out that both arbitrary spectral bandwidths and rotation angle values cannot be chosen independently and we discuss two configurations to overcome this constraint.

\section{Description of linear polarization rotation with total transmission}

\subsection{Theoretical background}

\begin{figure}[t]
	\centerline{\includegraphics[width=\textwidth]{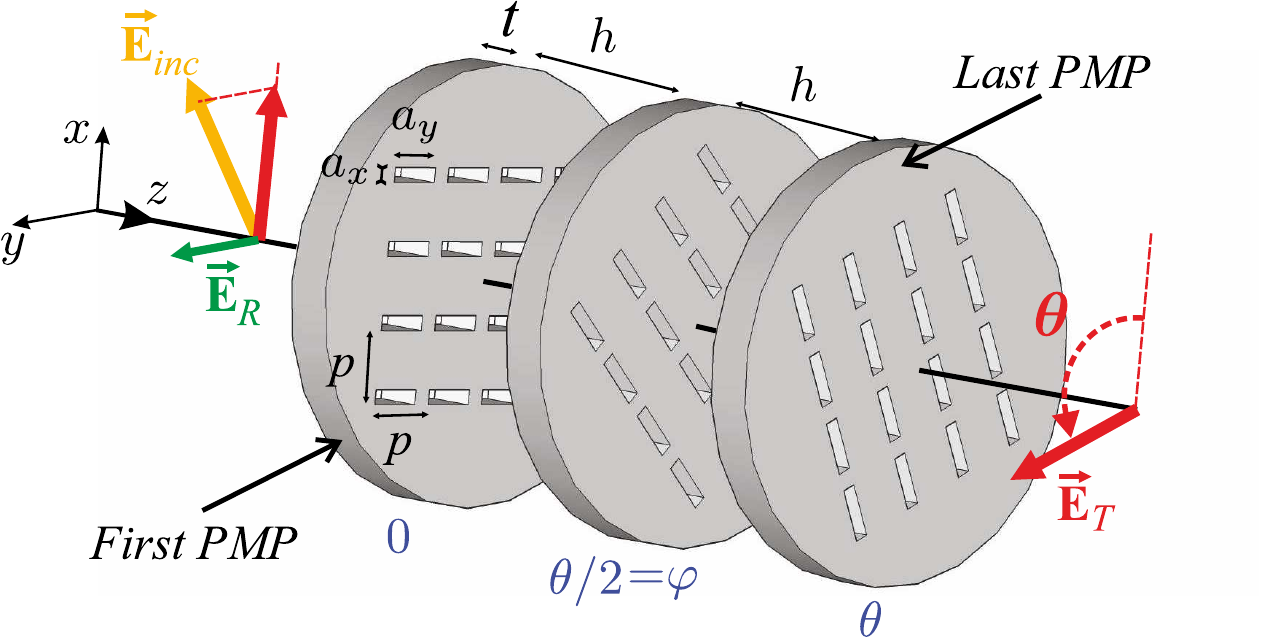}}
	\caption{\label{fig1} Principle of the polarization-selective rotation through a stack of PMP. The rectangle's length and width are $a_y$ and $a_x$ respectively ($a_y>a_x$), the period along $x$ and $y$ axes is $p$, the distance between two PMP is $h$ and their thickness is $t$. The angle $\theta$ corresponds to the axes difference between first and last polarizer and $\varphi$ is the intermediate angle between two polarizers.}
\end{figure}

We consider a multi-layered structure made of identical and parallel PMP separated by a distance $h$, as shown in Fig. \ref{fig1}. Each PMP consists of a periodic array of specific subwavelength apertures pierced in a metallic screen, and behaves as  a total linear polarizer of the electric field \cite{art:Boyer14}. In this paper, all dielectric and non-metallic regions are filled with air, and metal is assumed to be a perfect electric conductor. Our study involves PMP patterned with rectangular holes but the results shown in this paper can be readily extended for other particular holes cross-sections \cite{art:Boyer14}. In our case, $a_x$ is the rectangle's width, $a_y$ is the rectangle's length, $p$ is the period along the $x$ and $y $ axes and the PMP thickness is denoted by $t$. All geometrical notations are represented in Fig. \ref{fig1} and length units are not mentioned since no dispersion for dielectric materials is taken into account.We investigate stacks of $N$ PMP, with $N \in \mathbb{N}^*$, on which a plane wave falls at normal incidence with an arbitrary electric polarization state. We consider $N\geq 3$ in order to obtain polarization-selective rotation phenomena.

\begin{figure}[t]
	\centerline{\includegraphics[width=\textwidth]{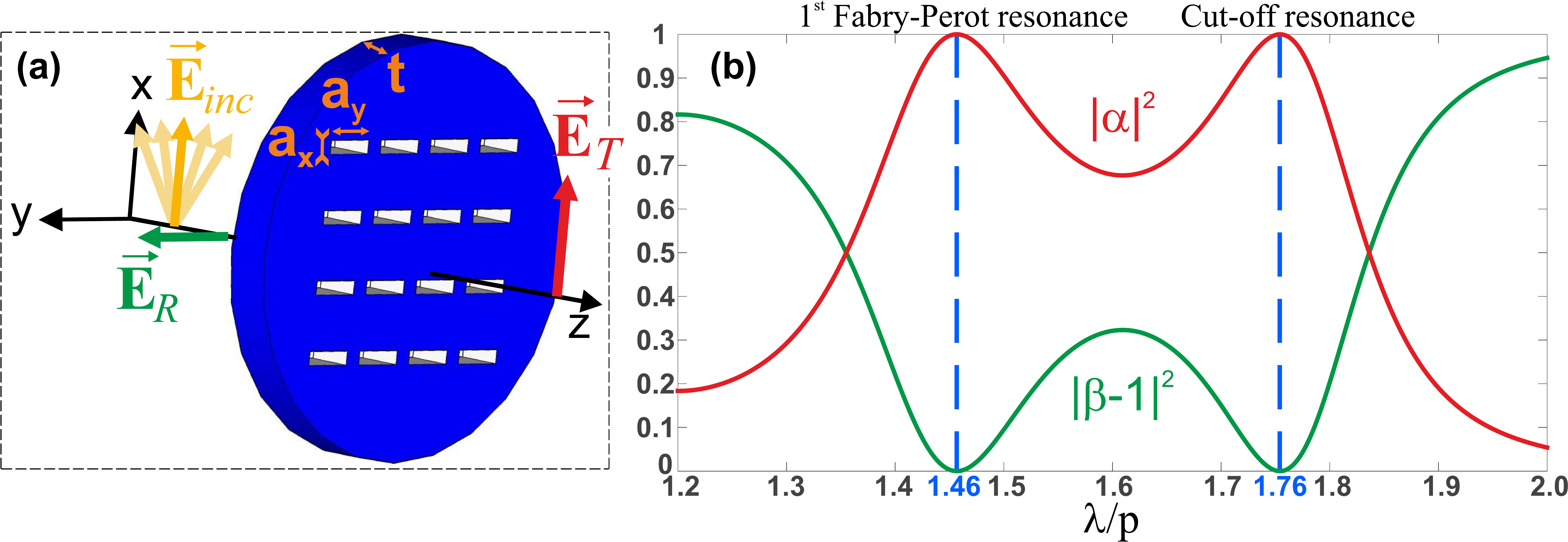}}
	\caption{\label{TvL} (a) Principle of a single polarization-selective PMP where the total transmitted and reflected electric fields, $\vec{E}_T$ and $\vec{E}_R$, are shown at resonance. The parameters are $a_x/p=0.3$, $a_y/p=0.9$, $t/p=1$. (b) $|\alpha|^2$ spectrum in red  and $|\beta - 1|^2$ spectrum in green  of a single PMP oriented along the x-axis. It reveals two resonances marked by the blue dotted lines that are located at $\lambda/p \approx 1.46$ and $\lambda/p \approx 1.76$ which respectively corresponds to the first Fabry-Perot resonance of the fundamental guided mode $TE_{01}$ and the cut-off resonance of this same mode.}
\end{figure}

For more clarity, let us first describe the special case where we consider a single PMP transmitting along the x-axis which means that we consider only the first PMP in Fig. \ref{fig1}. For this PMP, we have $\vec{E}_T = J^T\vec{E}_{inc}$ and $\vec{E}_R = J^R\vec{E}_{inc}$ where the transmission and reflection Jones matrices are
\begin{equation}
J^T=\left (\begin{array}{cc}
\alpha & 0 \\
0 & 0
\end{array}\right ) ; \ \ J^R=\left (\begin{array}{cc}
\beta - 1 & 0 \\
0 & -1
\end{array}\right )
\label{eq:JTRMono}
\end{equation}
and $\alpha$ and $\beta$ are resonant Airy-like terms in transmission and reflection respectively \cite{art:Boyer14}. Since we consider stacks of PMP with identical materials and geometry, the terms $\alpha$ and $\beta$ are the same for each PMP and  their expressions are given by \cite{art:Boyer14}
\begin{equation}
 \alpha=\dfrac{4\eta_{0}\eta u_t}{(C+\eta)^2-u_t^2(C-\eta)^2} \left |g_0\right |^2
 \end{equation}
and
\begin{equation}
\beta=\dfrac{2\eta_0\left [(C+\eta)+u_t^2(\eta - C)\right ]}{(C+\eta)^2-u_t^2(C-\eta)^2} \left |g_0\right |^2
\end{equation}
where $\eta_0$ is the relative admittance of the $0$th diffracted order in homogeneous regions and $\eta$ is the relative admittance of the cavity mode. C corresponds to the coupling coefficient of diffracted propagative and evanescent waves with the single fundamental mode $TE_{01}$ guided inside the apertures, and $u_t=\exp(ik_0t)$ is the propagation term of the mode inside the apertures. The term $g_0$ is the overlap integral between the 0th order diffracted wave and the $TE_{01}$ mode. A more detailed description is given in ref. \cite{art:Boyer14}. 

Figure. \ref{TvL}(a) shows the principle of linear polarization of one PMP where the electric fields $\vec{E}_T$ and $\vec{E}_R$ polarization are depicted at resonances. The parameters are $a_x/p=0.3$, $a_y/p=0.9$ and $t/p=1$. Figure \ref{TvL}(b) gives the corresponding $|\alpha|^2$ spectrum (red curve) and $|\beta - 1|^2$ spectrum (green curve) which let appear two resonances. The peak close to $\lambda/p=1.76$ is related to the cut-off of the $TE_{01}$  and the peak close to $\lambda/p=1.46$ refers to its first Fabry-Perot-like resonance \cite{art:Boutria12}.

In this paper, the principle of LPR consists in rotating the incident linear electric field $\vec{E}_{inc}$ component with respect to the x-axis by an angle $\theta$ after transmission through the last PMP, as shown in Fig. \ref{fig1}. Hence, the first PMP is oriented such that its transmitted electric electric field $\vec{E}_T$ is along the x axis (term $\alpha$ of $J^T$ in eq. \ref{eq:JTRMono}). In other words, the incident electric field component along the y-axis is not transmitted nor rotated but totally reflected by the first PMP and is denoted by $\vec{E}_R$ (term $-1$ of $J^R$ in eq. \ref{eq:JTRMono}). The rotation angle of each PMP axis uniformly changes from $0$ to $\theta$ in the transverse $Oxy$-plane. We introduce $\varphi$ as the uniform rotation angle between two successive PMP, so that
\begin{equation}
\left(N-1\right)\varphi=\theta \textrm{ \textit{mod} } \pi.
\label{eq:relThPhi}
\end{equation}

We have developed a simple theoretical model which illustrates the physical principle underlying the LPR effect. This present model is a generalized form to an arbitrary PMP number $N\geq 3$ of the theory previously used for two stacked PMP in ref. \cite{art:Romain16}. We analytically deduce the expressions of the transmission and reflection Jones matrices, denoted by $J^T$ and $J^R$ respectively, for a stack of $N$ PMP:
\begin{equation}
J^T=\left(\begin{array}{cc}J^T_{x,x} & 0\\J^T_{y,x} & 0\end{array}\right)
=\alpha_N\left(\cos\varphi\right)^{N-1}\left(\begin{array}{cc}\cos\theta & 0\\\sin\theta & 0\end{array}\right),
\label{eq:JT}
\end{equation}
and
\begin{equation}
J^R=\left(\begin{array}{cc}J^R_{x,x} & 0\\0 & J^R_{y,y}\end{array}\right)
=\beta_{N}\left(\begin{array}{cc}1 & 0\\0 & 0\end{array}\right)-I_d,
\label{eq:JR}
\end{equation}
where $I_d$ is the identity matrix. The terms $\alpha_{N}$ and $\beta_{N}$ correspond to the Airy-like spectral resonant coefficients for the whole structure. Their analytical expressions are obtained by using an iterative process that account for the PMP stacking. More details can be found in  \cite{art:Romain16,art:Li96}.  After tedious calculations, the coefficients $\alpha_{N}$ and $\beta_{N}$ in eqs. (\ref{eq:JT}) and (\ref{eq:JR}) are  
\begin{equation}
\alpha_{N}=\frac{\alpha_{N-1}\alpha u}{\gamma_{N}-u^2\left(1-\beta\right)\left(1-\beta_{N-1}\right)},
\label{eq:alphT}
\end{equation}
and
\begin{equation}
\beta_{N}=\alpha\frac{\alpha_{N}}{\alpha_{N-1}}\left(\beta_{N-1}\delta-1\right)u+\beta,
\label{eq:alphR}
\end{equation}
with
\begin{equation}
\gamma_{N}=\frac{1-u^2\left(1-\beta\beta_{N-1}\sin^2\varphi\right)}{1-u^2}
\textrm{, }
\delta=\frac{\cos^2\varphi-u}{1-u^2},
\end{equation}
where $u=\exp{(ik_0h)}$ is the propagation term in homogeneous layers. Equations (\ref{eq:alphT}) and (\ref{eq:alphR}) are iterative formulas where the initial terms $\alpha_1$ and $\beta_1$ correspond to $\alpha$ and $\beta$.

We remind that this theory is restricted to the far-field approximation and it implies that $h$ is large enough to neglect coupling between evanescent waves at interfaces of two successive PMP \cite{art:Romain16,art:Boyer14}. Thus, the transmission involves only propagating electromagnetic fields which are reduced to the 0th diffracted propagative order in the considered spectral range. Nonetheless, the evanescent waves in homogeneous regions are taken into account in the transmission and reflection properties of each PMP through the coupling coefficient C (see eq. (6) in \cite{art:Boyer14}). Besides, the wavelength validity domain of the theory is chosen such that only the $TE_{01}$ mode exists inside rectangular apertures.

\subsection{Total transmission of linear polarization rotation}

\begin{figure}[t]
	\centerline{\includegraphics[width=\textwidth]{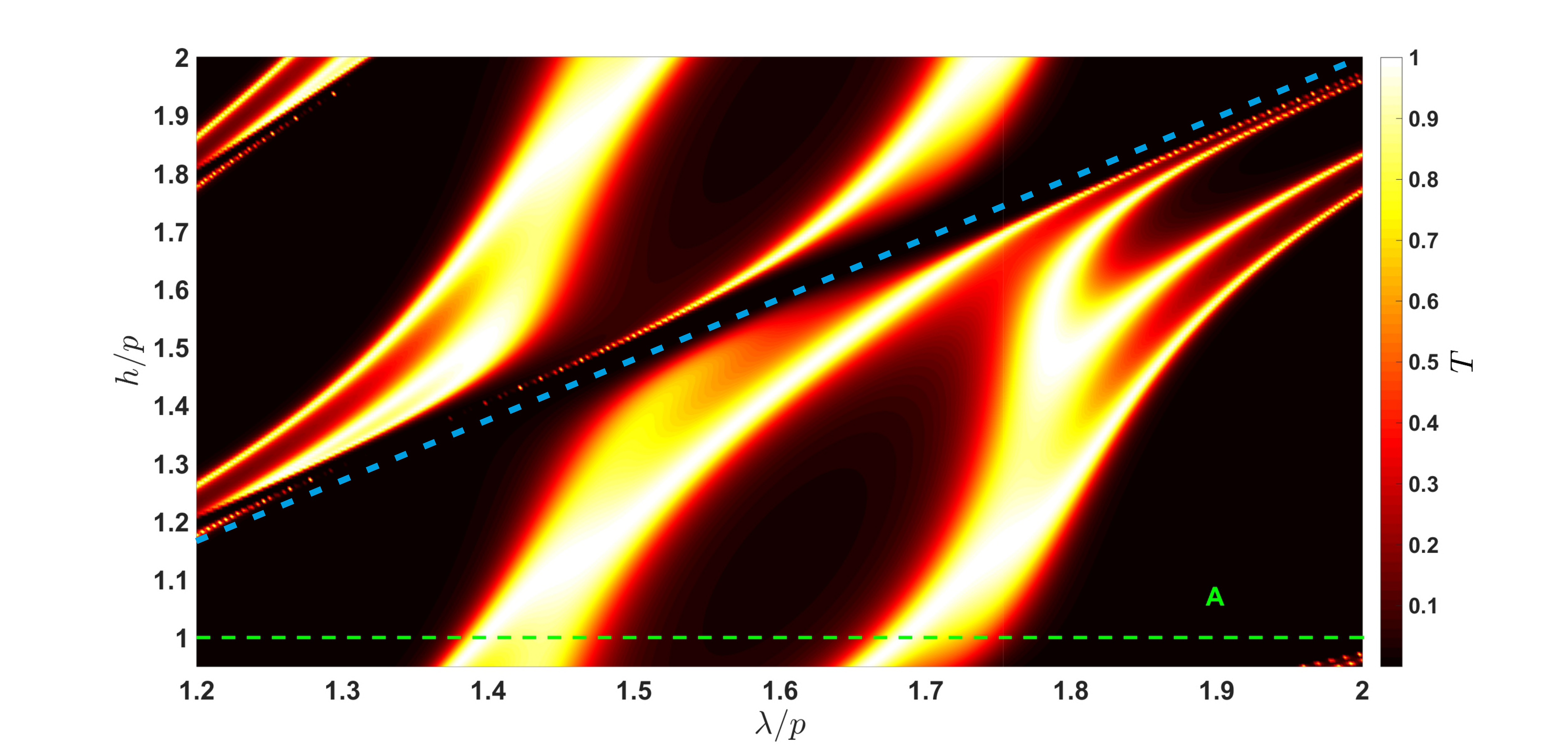}}
	\caption{\label{fig2lin} Transmission spectra for different values of $h/p$ for $N=3$. Other parameters are $a_x/p=0.3$, $a_y/p=0.9$, $t/p=1$ and $\theta=90^o$ ($\varphi=45^o$).}
\end{figure}

To study the transmission properties of our structure, we define the transmission T such that
\begin{equation}
T=\left|J^T_{x,x} \right|^2+\left|J^T_{y,x} \right|^2=\left|\alpha_{N}\left(\cos\varphi\right)^{N-1}\right|^2
\label{eq:PRR}
\end{equation}
It is important to notice that the transmission is an intrinsic property of the structure and do not depend on the incident light polarization. We first investigate the case $N=3$. The parameters are $a_x/p=0.3$, $a_y/p=0.9$, $t/p=1$ and $\varphi=45^{\circ}$ ($\theta=90^{\circ}$). Figure \ref{fig2lin} shows the transmission versus $\lambda/p$ and versus the distance between polarizers $h/p$ in linear and logarithmic scale respectively. The quantity $h/p$ is thereafter chosen equal to $1$, i.e such that, at least, two distinct resonant transmission peaks occur in $\lambda/p \in [1.2,2.0]$ (see green dotted line in Fig. \ref{fig2lin}. The LPR with such parameter values is labelled as $\textbf{A}$ in Fig. \ref{fig2lin}.

It may be surprising to get total transmission at resonances by simply interposing one linear polarizer at $45^{\circ}$ between two crossed ones. According to eq. (\ref{eq:JT}), this means that $T=\left|\alpha_{3}\right|^2\left[\cos\left(\pi/4\right)\right]^4=1$ at resonances. We expect that this total transmission is due to multiple reflections between PMPs \cite{art:Romain16}. For cascaded dichroic polarizers, where the multiple reflections vanish, $T=\left|\left(\cos\varphi\right)^{N-1}\right|^2$ which is equivalent to eq. (\ref{eq:PRR}) with $\left|\alpha_{N}\right|=1$ $\forall\lambda/p$. In this case and for $N=3$, the transmission reaches only $25\%$ for dichroic polarizers instead of $100\%$ at resonance of PMP, as shown in Fig. \ref{fig2lin}. This perfectly agrees with the classical Malus Law and demonstrates that the multiple reflections between PMP are responsible for the observed total transmission at resonance.

We may also expect that peak positions are largely dependent on the homogeneous layer's thickness $h$ because of these multiple reflections. However, the transmission spectra for different values of $h/p$ shown in Fig. \ref{fig2lin} exhibit peaks with positions that roughly coïncide with the ones of a single PMP ($\lambda/p \approx 1.4$ and $1.68$) except when the Fabry-Perot resonances of the cavity formed by two consecutive PMP intersect the PMP own resonances which are independent of $h$ (see oblique blue dotted line in Fig. \ref{fig2lin}). We believe that such a total transmission is a complex phenomenon that is out of the scope of this paper. Nevertheless, a more detailed analysis can be found in ref. \cite{art:Romain16}.

\section{Spectrally tunable linear polarization rotation}

\begin{figure}[t]
	\centering	
	\subfigure[]{
	\label{fig3a} 
	\includegraphics[width=0.49\textwidth]{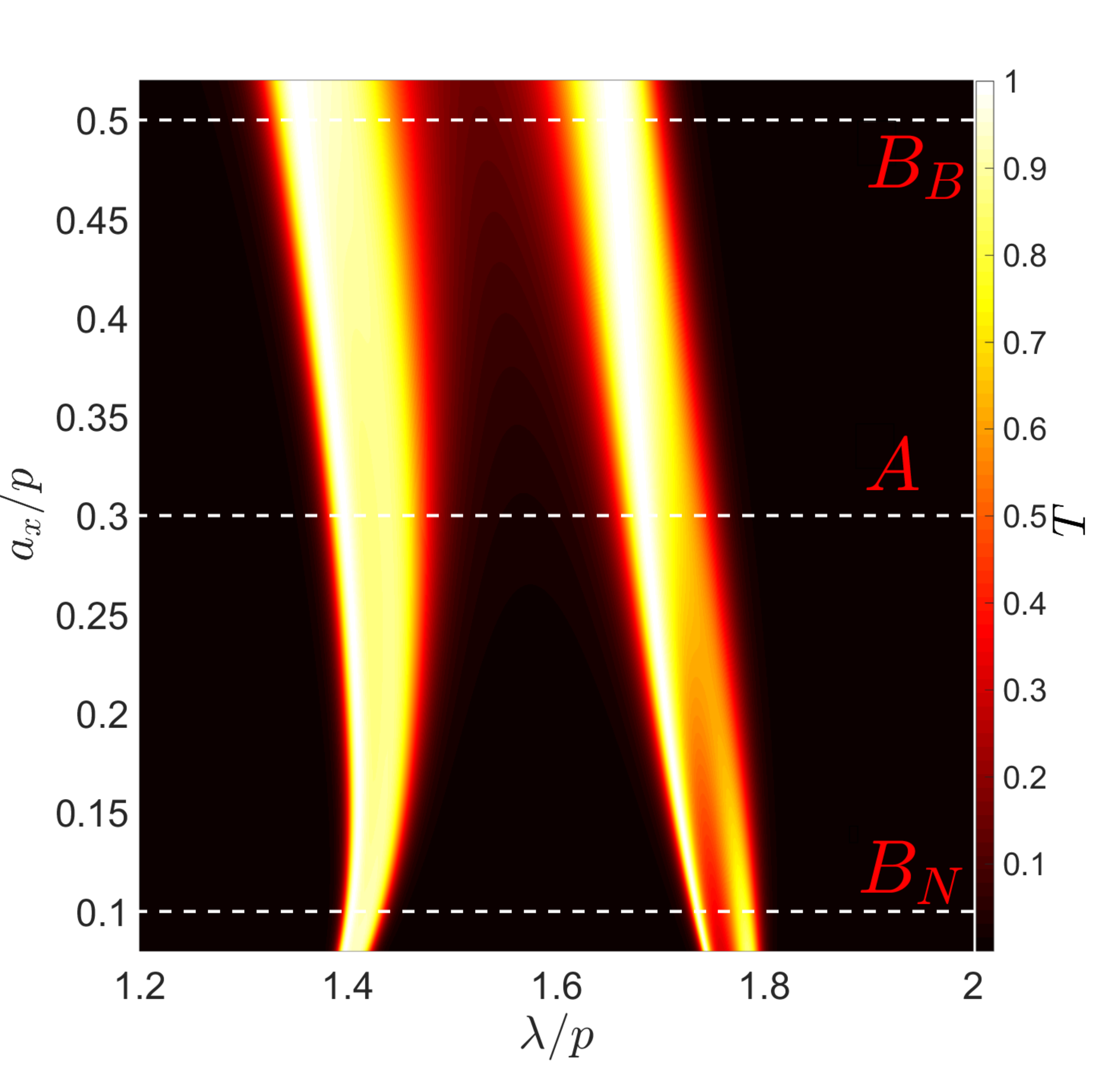}}
	%\hspace{0.2cm}
	\subfigure[]{
	\label{fig3b} 
	\includegraphics[width=0.48\textwidth]{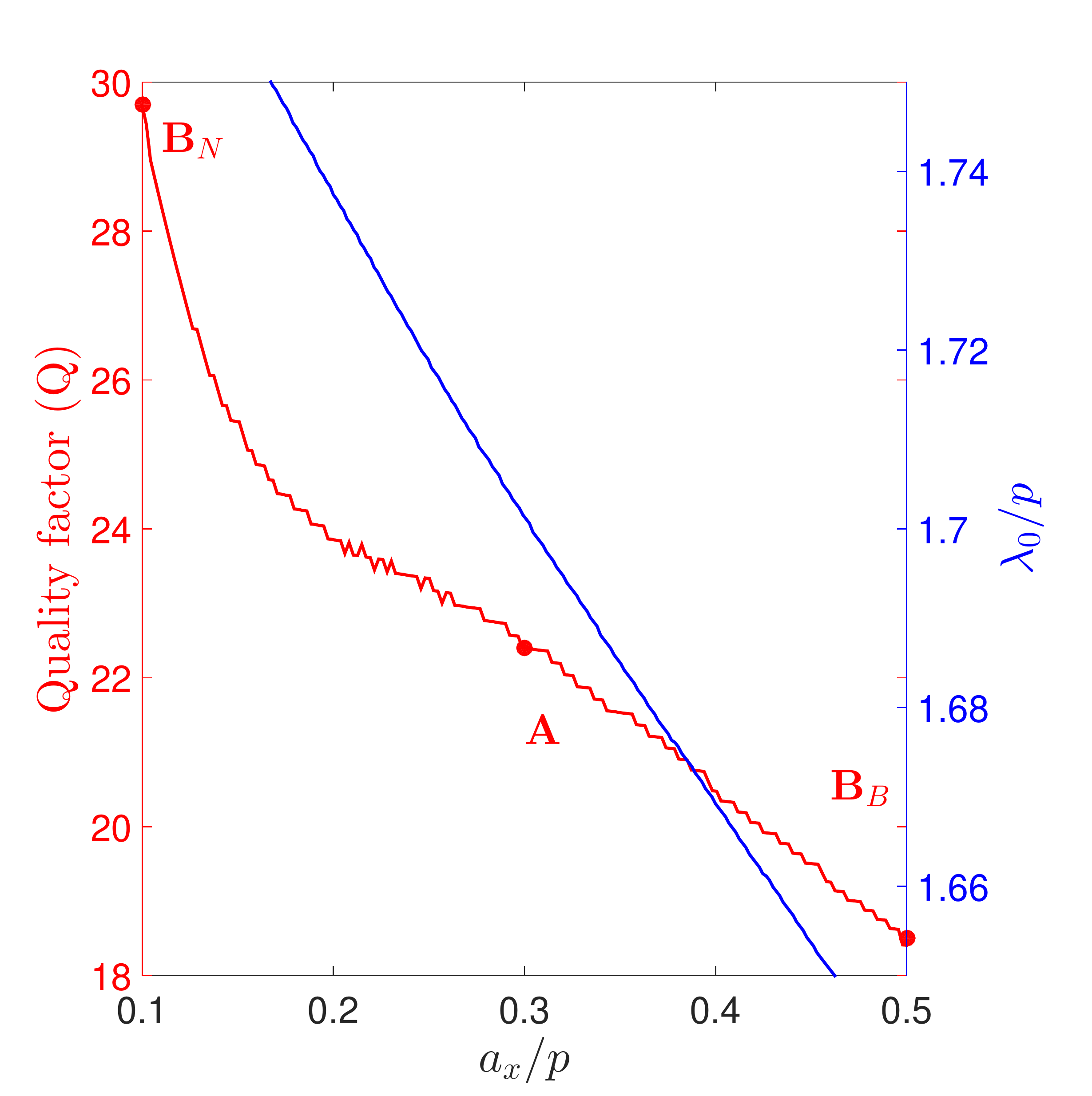}}
	\caption{(a) Transmission spectra for different values of $a_x/p$ for $N=3$. (b) Quality factor ($Q$) and spectral position ($\lambda_0/p$) of the peak close to $\lambda/p=1.68$. Other parameters are $a_y/p=0.9$, $t/p=h/p=1$ and $\theta=90^o$ ($\varphi=45^o$).}
\end{figure}

In this section, we show that the arrangement of the structure, as well as the holes geometry, are playing a crucial role to control the LPR quality factor. However, the rotated output field polarization direction, given by $\theta$, and the quality factor, which can be influenced by $\varphi$ cannot be arbitrarily and simultaneously fixed. Hence, we first focus on the angle $\varphi$ with no consideration for $\theta$.

From now on, we restrict our study to the peak close to $\lambda/p=1.68$ which is related to the cut-off of the $TE_{01}$ guided mode inside the rectangular holes. However, the cut-off wavelength of the $TE_{01}$ mode is equal to $2a_y$. This implies to keep the same value of $a_y$ in order to avoid important peak shifts. In section \ref{sec:BPR}, the study is led to design Broadband LPR (BPR) which corresponds to  low quality factors (Q$<10$). Then, we discuss the limitations of the BPR. Conversely, in section \ref{sec:NPR} we design a Narrowband LPR (NPR) which corresponds to ultra-high quality factors (Q$>10^5$).

\subsection{Broadband linear polarization rotation}
\label{sec:BPR}
First, $a_x/p$ must be chosen as large as possible to obtain BPR. Nevertheless, the value of $a_x/p$ must be chosen such that the cut-off frequency of the second mode $TE_{10}$ remains smaller than the first Rayleigh-Wood frequency. In our case, the maximum value of $a_x/p$ is $0.5$. Results shown in Fig. \ref{fig3a} confirm that the width of transmission peaks increases when $a_x/p$ grows. Precisely, Fig. \ref{fig3b} shows the quality factor $Q=\lambda_0/\Delta\lambda$ computed for the peak close to $\lambda/p=1.68$, for which $\lambda_0$ is the normalized central values of peaks and $\Delta\lambda$ corresponds to the FWHM. It particularly shows that Q decreases when the rectangle's width, $a_x/p$, increases. Thereafter, we fix $a_x/p=0.5$ for BPR. The BPR with such parameter values is labelled as $\textbf{B}_B$ in Fig. \ref{fig3b} with $Q=18.4$. 

\begin{figure}[t]
	\centering	
	\subfigure[]{
	\label{fig4a} 
	\includegraphics[width=0.48\textwidth]{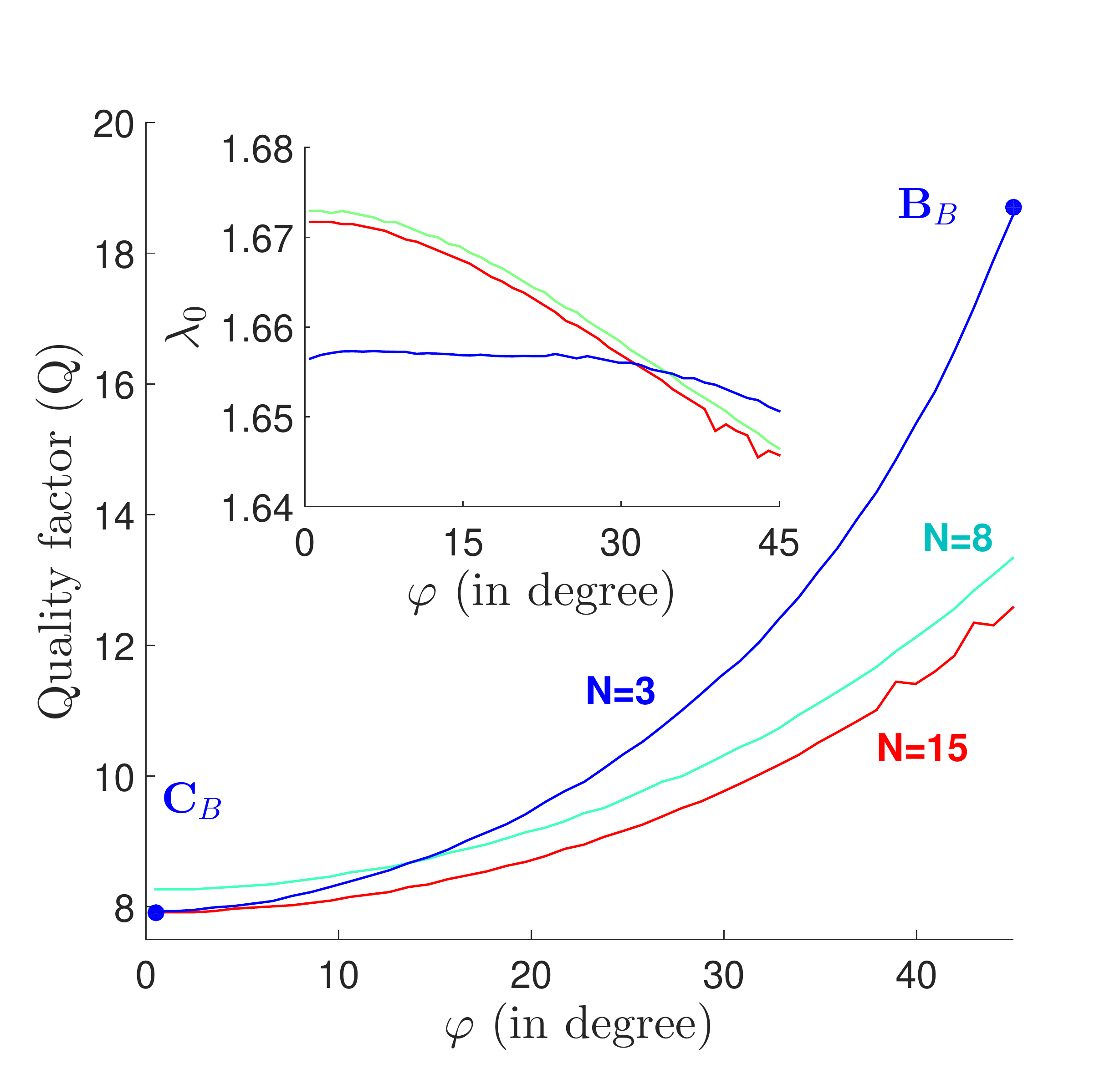}}
	\subfigure[]{
	\label{fig4b} 
	\includegraphics[width=0.48\textwidth]{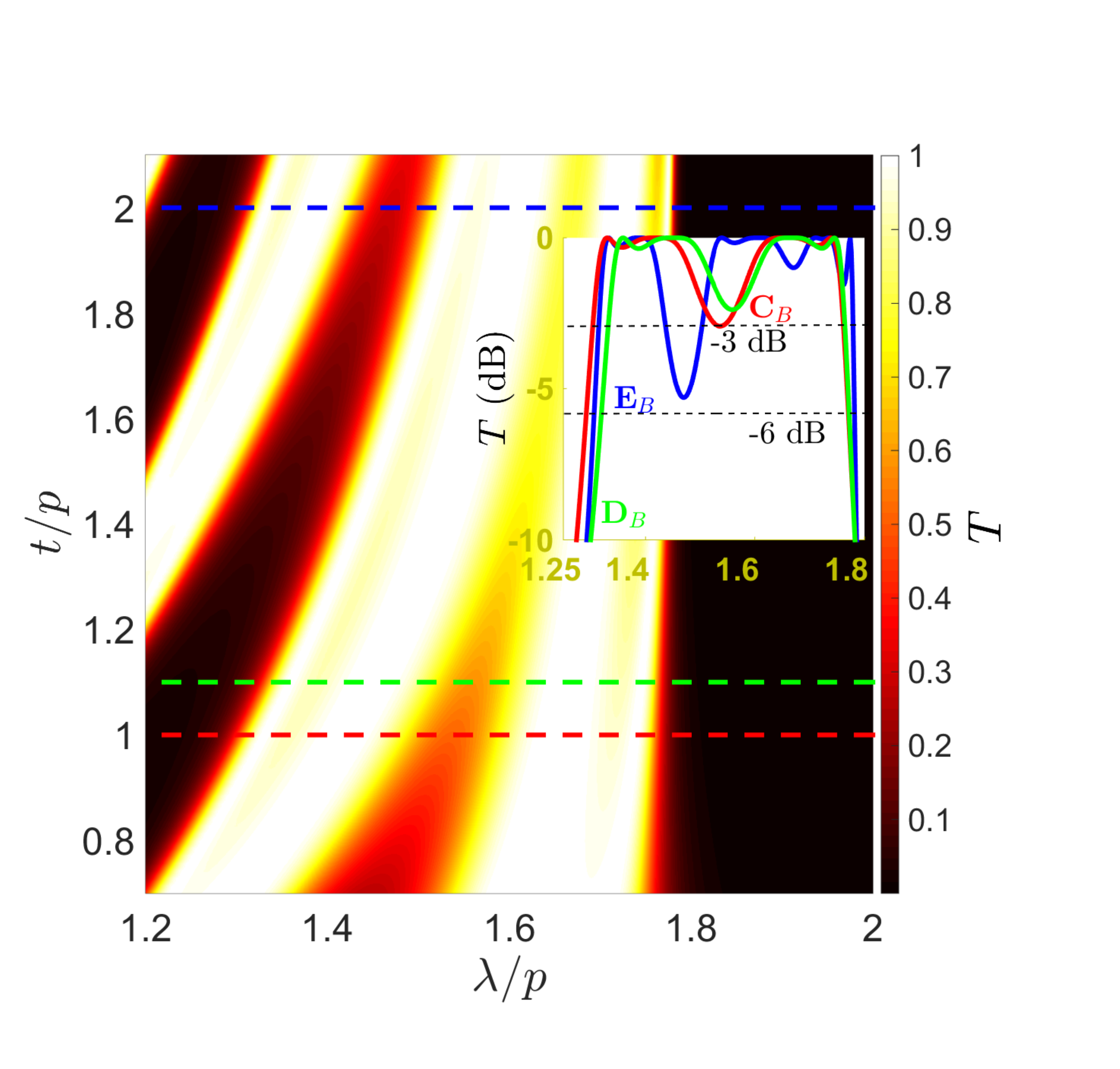}}
	\caption{(a) Variations of quality factor of the peak close to $\lambda/p=1.68$ with respect to $\varphi$ and for different values of $N$. Variations of the peak positions are shown in the inset graph. Other parameters are $a_x/p=0.5$, $a_y/p=0.9$ and $t/p=h/p=1$. (b) Transmission spectra for different values of $t/p$ for $N=3$. The transmission spectra for $t/p=1$, $t/p=1.1$ and $t/p=2$  are shown in the inset graph. Other parameters are $a_x/p=0.5$, $a_y/p=0.9$, $h/p=1$ and $\varphi=0.5^o$.}
\end{figure}

Figure \ref{fig4a} shows the variation of $Q$ with respect to $\varphi$ and for different values of $N$. As expected, the quality factor decreases when $\varphi$ decreases. Similarly to photonic crystal band gaps broadening, quality factors tend to one limit for a fixed value of $\varphi$ when the number $N$ of cascaded PMP increases. We also remark that a BPR with $N=3$ reaches identical performances at $\varphi=0.5^o$ than a BPR with $N=15$. The BPR, with $N=3$ and $\varphi=0.5^o$, is labelled as $\textbf{C}_B$ in Fig. \ref{fig4a} for which $Q=7.9$. The inset graph in Fig. \ref{fig4a} reveals that the peak position remains almost constant near $\lambda_0 \approx 1.65$.

Another idea to broaden the bandwidth is to merge the peaks by shifting the peak centered to $\lambda/p \approx 1.4$ to the one centered to $\lambda/p\approx 1.68$. The peak close to $\lambda/p=1.4$ is related to a Fabry-Perot-like resonance of the $TE_{01}$ cavity mode and its resonance wavelength changes with the metal thickness $t$. The position of the peak near $\lambda/p=1.68$, related to the cut-off of $TE_{01}$ cavity mode, is not affected by $t/p$ values because it only depends on $a_y$. Fig. \ref{fig4b} depicts transmission spectra for different values of $t/p$. As expected, we see that the two peaks merge but the bandwidth of each peak narrows when $t/p$ increases. The spectra shown in the inset of Fig. \ref{fig4b} computed for a BPR with $t/p=2$, and labelled as $\textbf{D}_B$, reveals a relatively low quality factor: $Q=3.29$. As we can observe in the inset spectra of Fig. \ref{fig4b}, it is possible to modulate the spectral bandwidth, or in other words, the quality factor Q, at a -3dB threshold while the bandwidth at -6dB is barely affected with the variation of $t$. Those results show that it is possible to lower the quality factor by increasing the rectangle's width, by reducing the angle $\varphi$ and by carefully choosing the layer's thickness $t$. In the same time, they demonstrate that Q converges to a limit value. These multilayered devices present transmission properties close to cascaded nanobars structure studied in \cite{art:Wang16}. It features similar broadband range with a very efficient LPR obtained by twisting. 

As mentioned above, the relation between $\varphi$ and $\theta$ given by eq. (\ref{eq:relThPhi}) does not allow to choose an accurate quality factor $Q$ and an arbitrary angle $\theta$ simultaneously. This issue is especially crucial for the achievement of a low $Q$ and tunable LPR. For example, in the case of a cross-polarization rotation where $\theta = 90^{\circ}$ and with $N = 3$, the angle $\varphi$ is necessarily equal to $45^{\circ}$ (see section 2. \ref{eq:relThPhi}). Such a value of $\varphi$ is not optimized to obtain a low $Q$ cross-polarization rotation. In this case, the corresponding quality factor is $Q = 11.3$ for $\lambda_0 \approx 1.68$. Thus, we discuss two options to overcome this limitation.

\begin{figure}[t]
	\centerline{\includegraphics[width=\textwidth]{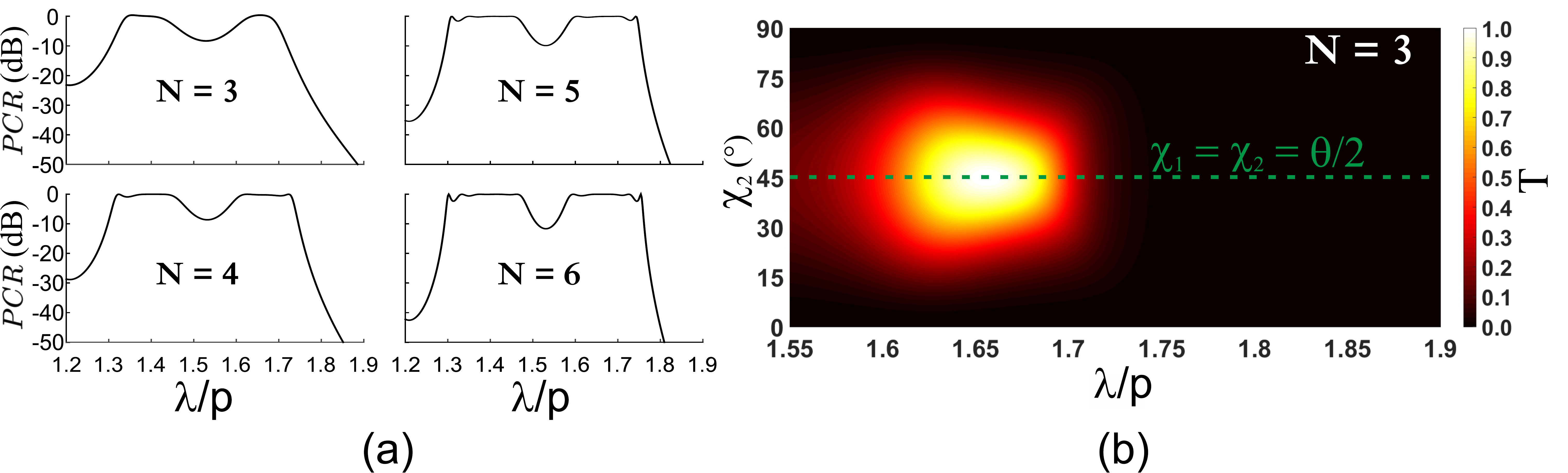}}
	\label{fig6}
	\caption{\label{fig6} (a) Transmission spectra for different values of $N$ for $\theta = 90^{\circ}$ with $a_x/p=0.5, a_y/p=0.9, t/p=h/p=1$. (b) Transmission spectra as a function of $\rchi_2$ the angle between the second and the third (last) PMP with $\rchi_1 + \rchi_2 = \theta$ and $\theta = 90^{\circ}$. }
\end{figure}

The first option simply consists in increasing $N$ in order to reduce $\varphi$ and therefore lower the quality factor. Figure \ref{fig6}(a) shows the transmission spectra for different values of $N$ for $\theta = 90 ^{\circ}$. As expected from section 3.1, we observe a broadening of the transmission spectra when $N$ increase. However, Fig. \ref{fig4a} and \ref{fig6}(a) shows that such a broadening is limited and converges to a finite spectral bandwidth (convergence of $Q$ when $\varphi \to 0^{\circ}$). Consequently, a reasonable number of polarizers should be chosen to achieve a good trade-off between a large bandwidth and a realistic device.

The second option is to consider a non-uniform angle $\rchi$ between each plate for a stack of $N$ polarizers such that $\sum\limits_{i=1}^{N-1} \rchi_i = \theta$ with $i \in \{1,2,\ldots , N-1\}$. Precisely, we study the simple case of $N = 3$ with $\rchi_1 + \rchi_2 = 90^{\circ}$, and Fig. \ref{fig6}(b) shows the transmission spectra as a function of $\rchi_2$. It is important to note that the transmission spectrum is optimized when $\rchi_1 = \rchi_2 = 45^{\circ}$ featuring a perfect transmission with the lowest Q. Thus, breaking the intermediate rotation angle $\varphi$ into different values $\rchi_i$ is not efficient for achieving a total and low-$Q$ LPR. Nevertheless, Fig. \ref{fig6}(b) also shows that the structure exhibits an angular tolerance within which its performances are barely affected. 

\subsection{Narrowband linear polarization rotation}
\label{sec:NPR}

\begin{figure}[t]
	\centerline{\includegraphics[width=\textwidth]{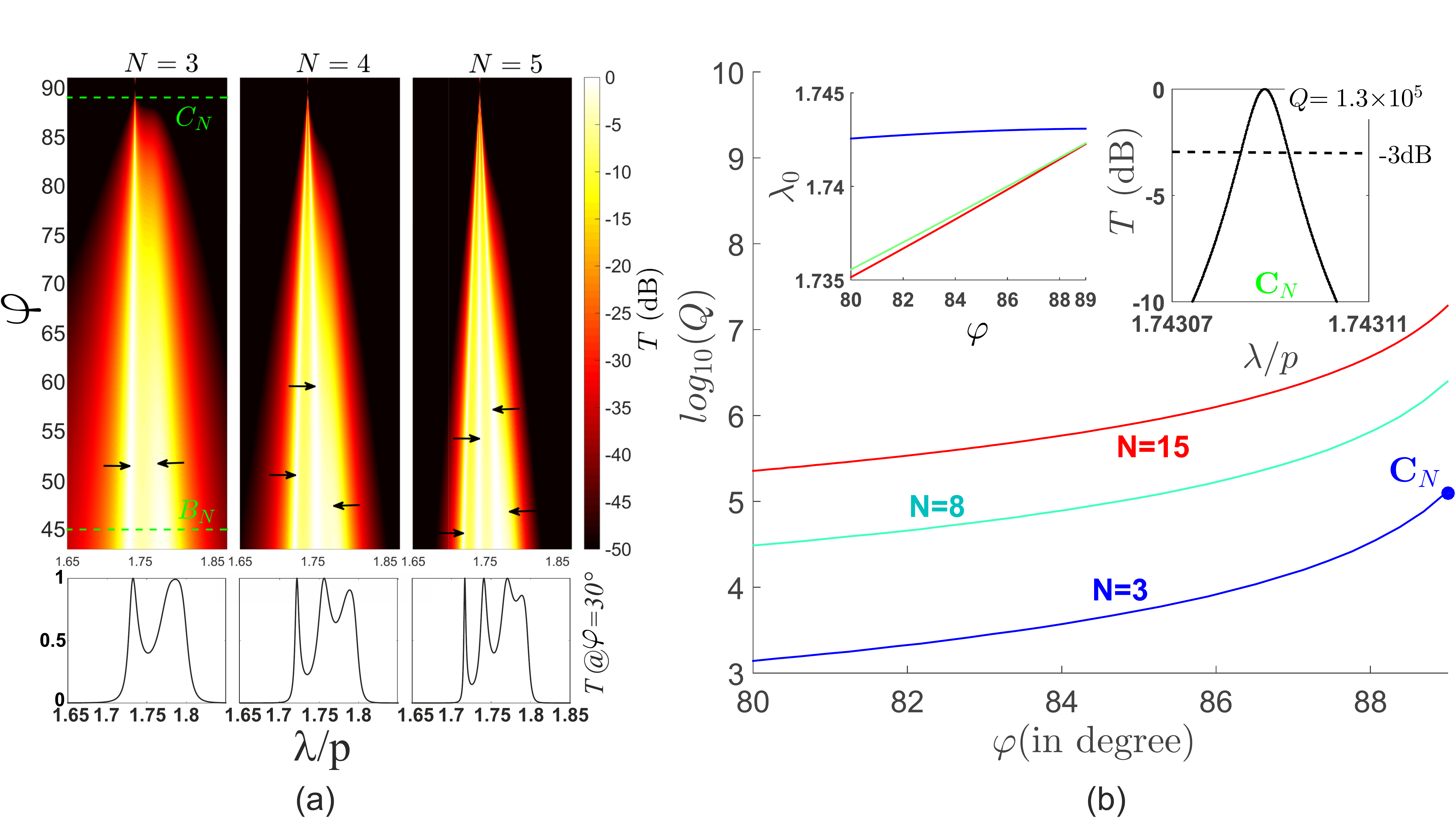}}
	
	\caption{\label{fig7}(a) Transmission spectra versus $\varphi$ for different values of $N$. (b) Variations of quality factor of the nearest peak to $\lambda/p=1.68$ in respect with $\varphi$ and for different values of $N$. Variations of the peak positions are shown in the inset graph on the left. Transmission spectra for $\varphi=89^o$ and $N=3$ is shown in the inset graph on the right. Other parameters are $a_x/p=0.1$, $a_y/p=0.9$ and $t/p=h/p=1$.}
\end{figure}

Contrary to BPR, $a_x/p$ must be chosen as small as possible in order to obtain NPR. Hence, the value of $a_x/p$ is fixed to $0.1$ which corresponds to the polarization rotation labelled as $\textbf{B}_N$ in Fig. \ref{fig3b}. For this value of $a_x/p$ and for $N=3$, the peak close to $\lambda/p=1.77$ shown in Fig. \ref{fig3a} splits into two peaks. The quality factor plotted in Fig. \ref{fig3b}, equal to $29.7$ at $\textbf{B}_N$, is computed for the global peak. We assume that this effect is due to resonance degeneracy. The whole structure may be seen as a stack of 5 cascaded and coupled resonators. Indeed, for $N=3$, there are 3 PMP resonators coupled to 2 multiple reflections resonances located in the two homogeneous regions. This complex behaviour deserves a thorough analysis in a future work. The narrow peaks are indicated by dark arrows in Fig. \ref{fig7}(a). Precisely, we see in graph on the left of Fig. \ref{fig7}(a) that the most narrow peak is increasingly being thin when $\varphi$ tends to $90^o$ while the other ones disappear. In general, $N-1$ peaks exist for $\varphi$ getting closer to $90^o$, one of which presents a relatively broad spectral bandwith with $T<1$, when the $N-2$ other ones are very narrow peaks with $T=1$ at resonances. In order to distinguish the peaks more easily, the peaks are also depicted in the lower inset spectra in Fig. \ref{fig7}(a) for $\varphi = 30^{\circ}$. It is interesting to note that, for fixed values of $N$, all narrow peaks seem to converge to a unique value of $\lambda/p$ when $\varphi$ increases. Thereafter, we study in particular the nearest peak to $\lambda/p=1.68$. 

We are now interested in the variation of the quality factor of the chosen transmission peak when $\varphi$ tends to $90^o$. The results are depicted in Fig. \ref{fig7}(b) and reveal that the quality factor drastically diverges when $\varphi$ tends to $90^{\circ}$ and for different values of $N$. As an example, the NPR with $\varphi=89^{\circ}$ and $N=3$ is labelled as $\textbf{C}_N$. For this case, $Q$ reaches $1.3\times 10^5$, and the transmission spectra is plotted in the inset graph on the right of Fig. \ref{fig7}(b). In the inset graph on the left, we remark that peak positions converge to a unique value of $\lambda/p$ when $\varphi$ increases and it confirms the observation made in Fig. \ref{fig7}(a). This interesting result could be used for the design of high quality filters for the THz spectral band. Nevertheless, in view of experimental demonstration, the robustness of the structure with respect to the fabrication imperfections should be discussed. However, the latter greatly depend on the  manufacturing process that is not already established. Such parametric study involving pure numerical methods, such as the finite difference time domain, must be applied to take into account the real geometry of the structure \cite{ndao:apl13}.

\section{Conclusion}
We have theoretically investigated a total LPR with an extremely tunable spectral bandwidth. More specifically, the polarizing devices consisted in stacks of PMP separated by dielectric layers. The numerical results are supported by a new and efficient model from which the Jones matrices of such stacked structures are analytically expressed. The optimized LPR is achieved by regularly rotating the successive PMP to mimic a chiral structure. Furthermore, we have underlined the influence of the rectangular holes' width and, more importantly, the angle $\varphi$ to selectively achieve a broadband or narrowband LPR. Precisely, we have shown that low quality factor ($Q < 10$) and high quality factor ($Q > 10^5$) can be obtained. Such flexible metadevices could be applied for the design of broadband linear cross-polarization rotators or high-Q filters, which represents important cornerstones in optical communications.

\end{document}